# Structures of ultrathin copper nanotubes

Jeong Won Kang* and Ho Jung Hwang

Semiconductor Process and Device Laboratory, Department of Electronic Engineering, Chung-Ang University, 221 HukSuk-Dong, DongJak-Ku, Seoul 156-756, Korea

We have performed atomistic simulations for helical multi-shell (HMS) Cu nanowires and nanotubes. Our investigation on HMS Cu nanowires and nanotubes has revealed some physical properties that were not dealt in previous works that considered metal nanowires. As the diameter of HMS nanowires increased, their cohesive energy per atom and optimum lattice constant decreased. As the diameter of HMS nanotubes increases, their cohesive energy per atom decreased but optimum lattice constant increased. Shell-shell or core-shell interactions mainly affected on the lattice constant and the diameter of HMS nanowires or nanotubes. This study showed that HMS nanotubes for materials of *fcc* metal crystals can be maintained when forces exerted on atoms of inner shell of the HMS nanotubes are zero or act on the direction of the outside.

Tel: 82-2-820-5296
Fax: 82-2-812-5318
E-mail: gardenriver@korea.com

# 1. Introduction

Recently, ultrathin metal nanowires have aroused growing interest in condensed matter physics; for example, Takayanagi's group has fabricated ultrathin gold [1-3] and platinum [4] nanowires. Many theoretical studies on ultrathin nanowires have been done using atomistic simulations for several metals, and these have simulated straight-line uniform ultrathin nanowires containing helical multi-shell (HMS) structures, such as Ag [5], Al [6], Au [7-10], Ti [11], Cu [12,13], and Zr [14]. Unlike a hexagonal network for carbon nanotubes, each shell of the HMS structures is formed by a triangular network which is similar to the {111} atomic sheet of *fcc* crystals. The <110> atomic rows in each {111} sheet make a helix that coils around the axis of metal nanowires. The $n$ - $n'$- $n''$- $n'''$ HMS nanowires, then, are composed of coaxial tubes with $n$, $n'$, $n''$, $n'''$ helical atom rows ($n > n' > n'' > n'''$).

In addition to studies on novel helical structures of ultrathin nanowires, the melting behavior of ultrathin nanowires has been investigated for Pb [15], Au [16], Cu [17], and Ti [18]. The compression of the HMS Au nanowires [19] and the tensile testing of the HMS Cu nanowires [20] have also been performed. The resonance of ultrathin Cu nanobridges was investigated using a classical molecular dynamics (MD) simulation [21]. In study on defects in the HMS Cu nanowires [22], the vacancy formation energy was lowest in the core of a HMS-type nanowire, a vacancy formed in the outer shell of a HMS-type nanowire naturally migrated toward the core, and an onion-like cluster with a hollow was also formed. These provided basic information on the formation of hollow HMS-type metal nanowires, and recently an evidence of a suspended a 13-6 HMS Pt nanotubes was reported [4].

Although the previous works have given support for metal nanowires of the HMS structures, further investigations, in areas such as non-linear ultrathin nanowires, funnel-shaped nanowires, defects in nanowires, and metal tubular structures like carbon nanotubes need to be made in order to understand the physical properties of nanowires, and for the successful application of nanowires to nanoscale devices. Therefore, this investigation focuses on copper HMS nanotubes and provides basic physical information on the structural properties of HMS-type nanotubes.

## 2. Computational methods

For the Cu-Cu interactions, we used a well fitted many-body potential function of the second-moment approximation of the tight-binding (SMA-TB) scheme [23]. This potential function reproduces many basic properties of crystalline and non-crystalline bulk phases and surfaces [24], and provides a good insight into the structure and thermodynamics of metal clusters [25,26]. Table 1 listed in Reference [27] shows that the physical values of Cu calculated by the SMA-TB method agree with other theoretical methods, and also with those measured by experiment. The SMA-TB potential has previously been used in atomic-scale simulation studies of nanoclusters [28-31] and nanowires [27,32-37]. We used the same values for parameters for the SMA-TB as those given in Reference [23]. The cut off distance, 5.30 Å, is the average distance between the fourth and fifth nearest-neighbours in a perfect crystal.

The optimum atomic arrangements were obtained using the steepest descent (SD) method, which is the simplest of the gradient methods, and so this was called the gradient descent method. The choice of direction was determined by where the force exerted by interatomic interaction decreased the fastest, which was in the opposite direction to $\nabla E_i$. In this work, the SD method was applied to the atomic positions, and the next atomic position vector ($r´_i$) was obtained by a small displacement of the existing atomic position vector ($r_i$) along a chosen direction under the condition, $|r´_i - r_i|/|\nabla E_i| = 0.001$.

Figure 1 shows a 5-1, a 6-1, and a 11-6-1 HMS Cu nanowires and a 6-0 and a 13-6 HMS Cu nanotubes, which are investigated in this paper. Each shell of the HMS nanowires and nanotubes was made by circular folding of {111} sheet and the HMS nanowires and nanotube were relaxed using the SD method. While the cores of the HMS nanowires were filled with atomic strand, the cores of the HMS nanotubes were empty. Each shell was composed of thirty atomic layers along the wire axis, and periodic boundary condition (PBC) was applied to supercells of nanowires and nanotubes. To provide easy understanding for structures of the HMS nanowires and nanotubes, we show the stripped structures of the 11-6-1 HMS nanowire and the 13-6 HMS nanotube in Fig. 1. Table 1 shows the number of atoms in supercells. The optimum lengths of PBCs of nanowires and nanotubes are related to the diameters of those, and this will be discussed in Table 1 and Fig. 2 in next chapter.

### 3. Results and discussion

We calculated optimum lattice constants of nanowires and nanotubes along the wire axis. Figure 2 shows the cohesive energy per atom as a function of lattice constant for the 5-1, the 6-1, and the 11-6-1 HMS Cu nanowires, and the points at the lowest cohesive energy are the optimum lattice constants. Using the structures of Fig. 1, as the lattice constants of those increased by 0.0001 Å from 2.22 Å, the optimum structure of each step was obtained from the SD method and then the cohesive energies per atom of those were calculated. As the number of shells in the HMS nanowires increases, their cohesive energy per atom and optimum lattice constant decrease. We also calculated the optimum lattice constants of nanotubes by using the same procedure, and results obtained are shown in Table 1.

Table 1 shows the optimum lattice constants, the cohesive energy per atom, and the mean diameter of shell composed of 6 atoms for the optimum structures of nanowires and nanotubes. As the diameter of the HMS nanowires increases, their lattice constants slightly decrease. The optimum lattice constant of the 6-0 nanotube is much smaller than those of the HMS nanowires. However, in the case of the 13-6 HMS nanotube, its lattice constant is higher than that of the 6-0 nanotube because of interaction between inner and outer shells, and is similar to those of the HMS nanowires. Therefore, this result appears that the shell - shell or core - shell interactions mainly affects on the lattice constant of the HMS nanowires or nanotubes. In the cases of the HMS nanowires, since their cores are filled with linearly atomic strand and their shells are made by circular folding of {111} sheets, both the distances between atoms in core and the heights of triangles in outer shell are mainly related to the lattice constants of the HMS nanowires. However, since the HMS nanotubes are only made by circular folding of {111} sheets, the heights of triangles are only related to the lattice constants of the HMS nanotubes. In Table 1, the lattice constant of the 6-0 nanotube is 93.33 % of that of the 6-1 nanowire. In the case of the 6-0 nanotube, if we assume that the length of a side of normal triangle be unit, 1, the height of triangle be $\sqrt{3}/2 = 0.866$. In the case of the 6-1 nanowire, if we assume that both the length of a side of normal triangle and the distance between atoms in core strand be unit, 1, the average between the height of triangle, $\sqrt{3}/2$, and the distance between atoms in core strand, 1, be $(2+\sqrt{3})/4 = 0.933$. Therefore, the lattice constant of the 6-0 nanotube is 92.82 % of that of the 6-1 nanowire from above two values. This value, 92.82 %, is in good agreement with 93.33 % obtained from our simulation. From these results, since the 6-1 nanowire has the atomic strand of core, the lattice constants and the diameters of the 6-1 nanowire are different with those of the 6-0 nanotube. In the case of double-shell nanotube, the 13-6 nanotube, the interaction between inner and outer shells also makes the longer lattice constant than a single-shell nanotube.

In previous works, the cohesive energies per atom ($E_{coh}$) for nanowires have been linearly proportional with the reciprocal of diameter ($D$) [6,33], and are expressed as follows,

$$E_{coh} \approx E_{bulk} + n/D, \qquad (3\text{-}1)$$

where $E_{bulk}$ is the cohesive energy of atom of bulk material and $n$ is a constant. In this investigation, since we considered only the 6-0 and the 13-6 HMS nanotubes, this paper could not provide relationship between the $E_{coh}$ and the $D$ for HMS nanotubes. However, our results show that the $E_{coh}$ of the 13-6 HMS nanotube with a double-shell is lower than that of the 6-0 nanotube with a single-shell. While the number of atoms of the 5-1 nanowire is equal to that of the 6-0 nanotube, the $E_{coh}$ of the 5-1 nanowire is lower than the $E_{coh}$ of the 6-0 nanotube. Our classical molecular dynamics simulations of the 6-0 nanotube frequently showed that the 6-0 nanotube was transformed into complex structures including the structure of a 5-1 nanowire. Therefore, we can insist that a 5-1 nanowire is more stable than a 6-0 nanotube.

We calculated the mean diameter ($D_{6s}$) of shell composed of 6 atoms for the 6-1 and 11-6-1 HMS nanowires and the 6-0 and 13-6 HMS nanotubes. In the case of the 11-6-1 HMS nanowire, since the outer shell slightly compresses the inner shell, the $D_{6s}$ of the 11-6-1 HMS nanowire is slightly smaller than the $D_{6s}$ of the 6-1 nanowire. The $D_{6s}$ of the HMS nanotubes is larger than the $D_{6s}$ of the HMS nanowires. While the lattice constant of the 6-1 nanowire is larger than that of the 6-0 nanotube, the $D_{6s}$ of the 6-1 nanowire is shorter than $D_{6s}$ of the 6-0 nanotube. Therefore, the difference between volumes of the 6-1 nanowire and the 6-0 nanotube is 2.296 %, and this result implies that a 6-0 nanotube is a different geometry of shell composed of 6 atoms in the condition of constant volume.

In previous work [22], as the diameter of HMS nanowire increased, the vacancy formation energy of its core decreased rapidly. MD simulations also showed that vacancy migrated from the outer shell to the inner shell and to the core. Since the formation energy of a vacancy was lowest at the core and the vacancy migrated towards the lower energy state, the vacancy migrated to the core. Therefore, these results implied that vacancies would be most frequently found in the core of a HMS nanowire. This interpretation is in good agreement with previous result that showed an evidence of metal nanotube and HRTEM images of a 6-0 and a 13-6 HMS Pt nanotubes. Therefore, we investigated an 11-6-1 HMS nanowire with a hollow region as shown in Fig. 3(a). Some core atoms in the center region of the 11-6-1 HMS nanowire was omitted and then this structure, Fig. 3(a), was relaxed by the SD method and transformed as Fig. 3(b). The region with a hollow of the 11-6-1 HMS nanowire was compressed as shown in Fig. 3(b). In this work, the forces exerted on atoms of inner shell were calculated, and the average value and direction of those was 0.717 eV/Å and toward the core, respectively. Figure 3(c) shows the side and cross-sectional views of region without an atomic strand of core. This result shows that forces exerted on atoms of inner or outer shell acted on the direction of the inside and the region with a hollow was compressed. Therefore, finally, the hollow region as shown in Fig. 3(a) disappeared as shwon in Figs. 3(b) and 3(c). On the analogy of this result, if forces exerted on atoms of shells of a HMS nanotube are zero or act on the direction of the outside, the nanotube will be a stable structure. This interpretation is also related to the mean diameter ($D_{6s}$) of shell composed of 6 atoms. As shown in Table 1, the $D_{6s}$ of nanotubes is larger than the $D_{6s}$ of nanowires. Especially, The $D_{6s}$ of the 13-6 HMS nanotube have the largest in this work, and this is because the interaction between inner and outer shells is attractive. Takayanagi group also showed the HRTEM image of a 13-6 Pt HMS nanotube obtained from a suspended nanowire made by electron-beam thinning method [4].

## 4. Conclusion

This study on HMS Cu nanowires and nanotubes has revealed some physical properties that were not dealt in previous works that considered metal nanowires. As the diameter of nanowires increased, their cohesive energy per atom and optimum lattice constant decreased. Shell-shell or core-shell interactions mainly affected on the lattice constant and diameter of HMS nanowires or nanotubes. Simulation result of a 11-6-1 HMS nanowire with a hollow region showed that forces exerted on atoms of inner or outer shell acted on the direction of the inside and finally, the region with a hollow was compressed. From this study and previous work reporting a evidence of HMS Pt nanotube [4], we conclude as follows: for materials of *fcc* metal crystals, when forces exerted on atoms of inner shell of HMS nanotubes are zero or act on the direction of the outside, the HMS nanotubes can be maintained.

**Table**

Table 1. For structures of HMS Cu nanowires and nanotubes, the number of atoms in supercell, the optimum lattice constant along the wire axis ($a$), the cohesive energy per atom ($E_{coh}$), and the mean diameter of shell composed of 6 atoms ($D_{6s}$).

| Structure | Number of atoms in supercell | $a$ (Å) | $E_{coh}$ (eV) | $D_{6s}$ (Å) |
|---|---|---|---|---|
| 5-1 nanowire | 180 | 2.2467 | –2.90616 | - |
| 6-1 nanowire | 210 | 2.2401 | –2.94531 | 4.78844 |
| 11-6-1 nanowire | 540 | 2.2396 | –3.51202 | 4.734468 |
| 6-0 nanotube | 180 | 2.0913 | –2.58686 | 5.012518 |
| 13-6 nanotube | 570 | 2.2184 | –2.98825 | 5.405746 |

**Figures**

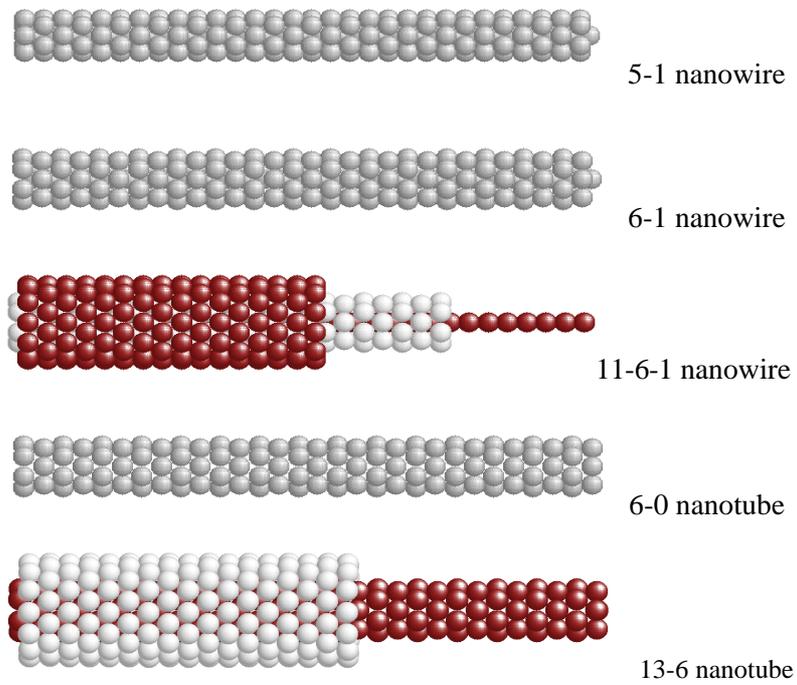

5-1 nanowire

6-1 nanowire

11-6-1 nanowire

6-0 nanotube

13-6 nanotube

Figure 1. Structures of well defined ultrathin copper nanowires and nanotubes obtained from the SD simulations at T = 0 K.

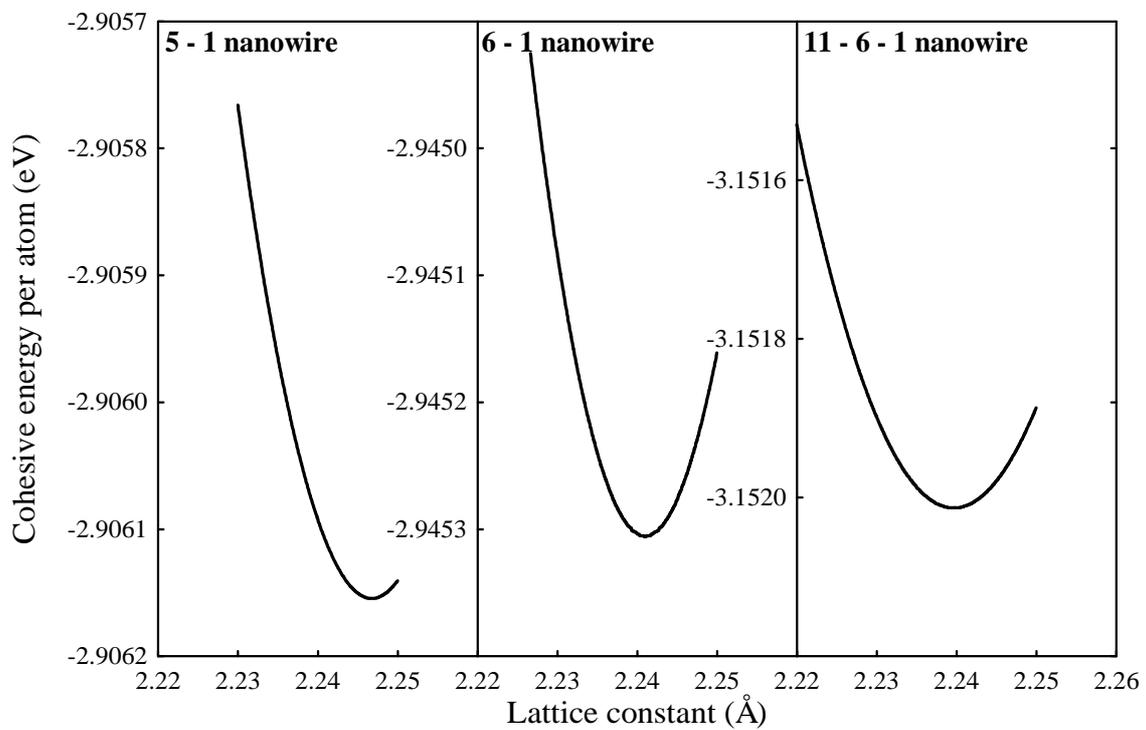

Figure 2. Total cohesive energy as a function of lattice constant for 5-1, 6-1, and 11-6-1 HMS Cu nanowires.

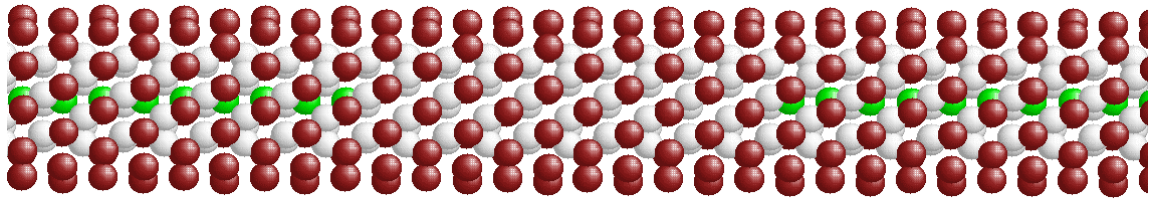

(a)

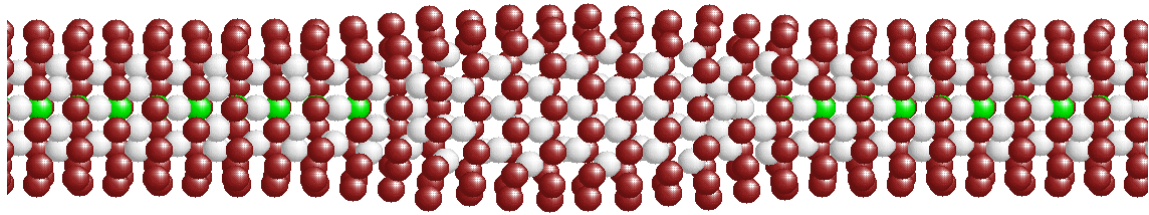

(b)

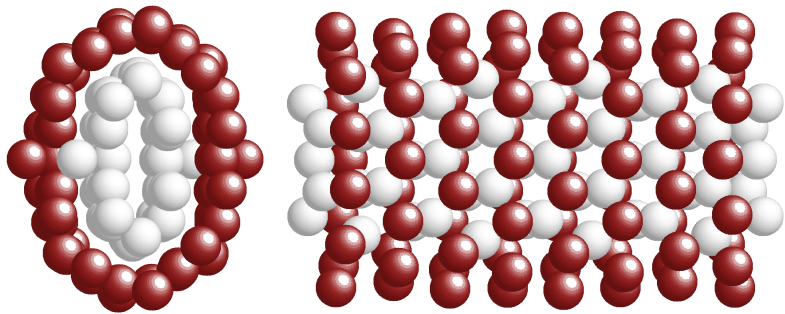

(c)

Figure 3. (a) Initial structure of 11-6-1 HMS Cu nanowire with a hollow, (b) structure relaxed by the SD method, and (c) side and cross-sectional views of region without an atomic strand of core in (b).